# Explosive instability due to 4-wave mixing

Benjamin R. Safdi and Harvey Segur
Department of Applied Mathematics, University of Colorado, Boulder, CO 80309-0526
(April 12, 2007)

*Abstract:* It is known that an explosive instability can occur when nonlinear waves propagate in certain media that admit 3-wave mixing. The purpose of this paper is to show that explosive instabilities can occur even in media that admit no 3-wave mixing. Instead, the instability is caused by 4-wave mixing: four resonantly interacting wavetrains gain energy from a background, and all blow up in a finite time. Unlike singularities associated with self-focussing, these singularities can occur with no spatial structure – the waves blow up everywhere in space, simultaneously.

PACS numbers: 05.45.-a, 42.65.-k, 47.35.-i, 52.35.-g

Among mathematical models that describe nonlinear wave propagation without dissipation, certain "universal" models stand out – each of these models appears when one takes a specific limit, and each arises in many physical situations. In all cases one first linearizes the governing equations about some trivial state, and obtains a (linearized) dispersion relation, $\omega(k)$, which relates the frequency ($\omega$) of a signal to its wavenumber ($k$). A 3-wave resonance is possible if the dispersion relation admits pairs $\{\omega_m, k_m\}$ such that

$$k_1 \pm k_2 \pm k_3 = 0, \qquad \omega_1 \pm \omega_2 \pm \omega_3 = 0. \tag{1}$$

For a given problem, (1) may or may not be possible. For example, in nonlinear optics, (1) occurs only in so-called $\chi_2$ materials [1]; for surface water waves, (1) is impossible for pure gravity waves, but it occurs if both gravity and surface tension are included in the model [2]. When (1) occurs, then $\{A_1(x,t), A_2(x,t), A_3(x,t)\}$, the slowly-varying complex amplitudes of three wave modes, evolve according to the "three-wave equations": three coupled equations of the form (with $l,m,n = 1,2,3$ cyclically)

$$\partial_t A_m + c_m \cdot \nabla A_m = i\delta_m A_n^* A_l^*. \tag{2}$$

Here $c_m$ is the group velocity and $\delta_m$ is a real-valued interaction coefficient, each corresponding to $\{\omega_m, k_m\}$ [3].

In the simplest model of 3-wave mixing, one ignores the spatial dependence of the interacting modes, so that (2) reduces to three coupled, complex, ordinary differential equations (ODEs),

$$\frac{dA_1}{dt} = i\delta_1 A_2^* A_3^*, \quad \frac{dA_2}{dt} = i\delta_2 A_3^* A_1^*, \quad \frac{dA_3}{dt} = i\delta_3 A_1^* A_2^*. \tag{3}$$

If any two $\delta_m$ in (3) differ in sign, then one can show that all solutions of (3) are bounded for all time. But this is not the only possibility: situations in which $\{\delta_1, \delta_2, \delta_3\}$ all have the same sign occur in plasmas [4,5], in density-stratified shear flows [6,7] and for vorticity waves [8]. If all $\delta_m$ have the same sign, then solutions of (3) can blow up in finite time (like $(t - t_0)^{-1}$), including solutions that start with arbitrarily small initial data. This is the *explosive instability*. All three waves grow together, so all three waves draw energy from a background source and blow up in unison. Thus, the relative signs of the $\delta_m$ in (2) and (3) signal whether such an energy source is available in the physical problem that (2) and (3) approximate.

The main point of this paper is to show that explosive instabilities can occur even in situations where a 3-wave resonance is impossible. In that case the simplest nonlinear interaction among wave modes is a 4-wave resonance, in which four pairs $\{\omega_m, \boldsymbol{k}_m\}$ satisfy

$$\boldsymbol{k}_1 \pm \boldsymbol{k}_2 \pm \boldsymbol{k}_3 \pm \boldsymbol{k}_4 = 0, \qquad \omega_1 \pm \omega_2 \pm \omega_3 \pm \omega_4 = 0. \qquad (4)$$

A common special case, in which one wave mode interacts with two other modes at nearly the same frequency and wave number,

$$(\boldsymbol{k} + \delta\boldsymbol{k}) + (\boldsymbol{k} - \delta\boldsymbol{k}) - \boldsymbol{k} = \boldsymbol{k}, \qquad (\omega + \delta\omega) + (\omega - \delta\omega) - \omega = \omega,$$

leads to the nonlinear Schrödinger (NLS) equation for the slowly varying, complex amplitude of one wave mode [3]

$$i\partial_t A + \{\alpha_1 \partial_x^2 + \alpha_2 \partial_y^2 + \alpha_3 \partial_z^2\} A + \gamma |A|^2 A = 0. \qquad (5)$$

Here $\{\alpha_m\}$ are real-valued constants obtained from $\omega(\boldsymbol{k})$, and $\gamma$ is a real-valued interaction coefficient (provided the original problem has no dissipation). In optics, (4) occurs in $\chi_3$ materials [1]. With one additional term, (5) becomes the Gross-Pitaevski equation, a commonly used model for Bose-Einstein condensates [9,10].

More complicated interactions, in which wave modes interact nonlinearly with themselves and also with other wave modes, lead to coupled NLS equations [11]. More complicated still are systems with self-interactions, cross-interactions, and 4-wave-mixing (with $m,p,q,r = 1,2,3,4$ cyclically):

$$i(\partial_t A_m + \boldsymbol{c}_m \cdot \nabla A_m) + \sum_{l,n} \alpha_{m,l,n} \partial_{x_l} \partial_{x_n} A_m + A_m \sum_{n=1}^{4} \gamma_{m,n} |A_n|^2 + \delta_m A_p^* A_q^* A_r^* = 0. \qquad (6)$$

The system in (6) has four such equations. In each equation, $\boldsymbol{c}_m$ is the group velocity and $\{\alpha_{m,l,n}\}$ are real-valued constants, all obtained from $\omega(\boldsymbol{k})$; $\{\gamma_{mn}\}$ are coefficients of NLS-type interaction terms; and $\{\delta_m\}$ are real-valued coefficients of the 4-wave mixing terms. The general form of this system of equations was first recognized in [12].

In this paper, we show that an explosive instability of the kind usually associated with 3-wave interactions can also occur because of 4-wave mixing. As with (3), a simpler model with 4-wave mixing than (6) is obtained by ignoring any spatial dependence of the interacting modes, so that (6) reduces to four coupled ODEs (with $m,p,q,r = 1,2,3,4$ cyclically):

$$i\frac{dA_m}{dt} + A_m \sum_{n=1}^{4} \gamma_{m,n} |A_n|^2 + \delta_m A_p^* A_q^* A_r^* = 0. \tag{7}$$

Note that with no spatial dependence, the self-focussing kind of singularity usually associated with NLS-type systems [13] cannot occur.

Without the 4-wave mixing terms in (7), there is no blow-up: one shows directly from (7) that for each $m$, if $\delta_m = 0$ then $|A_m|^2$ is constant. Hence we now assume that all $\delta_m \neq 0$. Then (7) admits three independent constants of the motion in the form of Manley-Rowe [14] relations:

$$J_m = \frac{|A_m|^2}{\delta_m} - \frac{|A_4|^2}{\delta_4}, \qquad m = 1,2,3. \tag{8}$$

It follows from (8) that if any two $\delta_m$ differ in sign, then every $|A_m|^2$ is bounded for all time. This result parallels the corresponding result for (3): all solutions of (3) are bounded if any two $\delta_m$ in (3) differ in sign. However, requiring that all the $\delta_m$ have the same sign in (7) is necessary but not sufficient for an explosive instability: it is also necessary that the $\delta_m$ be large enough relative to $\sum \gamma_{m,n}$, as we show next.

If all $\delta_m$ in (7) have the same sign, then change variables $\{A_m(t) = \sqrt{|\delta_m|} \cdot B_m(t)$, $\Gamma_{m,n} = \gamma_{m,n} |\delta_n|$, $\delta = sign(\delta_1)\sqrt{\delta_1 \delta_2 \delta_3 \delta_4}\}$ to obtain an equivalent system of ODEs ($m,p,q,r = 1,2,3,4$ cyclically):

$$i\frac{dB_m}{dt} + B_m \sum_{n=1}^{4} \Gamma_{m,n} |B_n|^2 + \delta \cdot B_p^* B_q^* B_r^* = 0. \tag{9}$$

This system of ODEs is Hamiltonian, with conjugate variables $\{B_m, B_m^*, m = 1,2,3,4\}$ and Hamiltonian $H = H_1 + H_2$, where

$$H_1 = -\frac{i}{2} \sum_{m,n=1}^{4} \Gamma_{m,n} B_m B_m^* B_n B_n^*, \quad H_2 = -i\delta(B_1 B_2 B_3 B_4 + B_1^* B_2^* B_3^* B_4^*). \tag{10}$$

Direct computation shows that $H$ is a constant of the motion. In addition, in these variables, (8) becomes

$$J_m = |B_m|^2 - |B_4|^2, \qquad m = 1,2,3. \tag{8a}$$

And one can verify directly that the usual Poisson bracket of any two of $(H, J_1, J_2, J_3)$ vanishes, so these constants are said to be in involution. Then it follows that the system of four complex ODEs in (9) is completely integrable in the sense of Liouville [15].

Next, we show that solutions of (9) blow up in finite time if

$$|\sum_{m,n=1}^{4}\Gamma_{m,n}| \leq 4|\delta|. \tag{11a}$$

In terms of the variables in (7), its solutions blow up in finite time if all four $\delta_m$ have the same sign and

$$|\sum_{m,n=1}^{4}\gamma_{m,n}|\delta_n|| \leq 4\sqrt{\delta_1\delta_2\delta_3\delta_4}. \tag{11b}$$

In either notation, these are the criteria for an explosive instability due to 4-wave mixing. They comprise the main result in this paper. Assuming (11) holds, a four-parameter family of exact, singular solutions of (9) is

$$B_m(t) = \frac{c \cdot e^{i\theta_m}}{(t_0 - t)^{\frac{1}{2} + i\phi_m}}, \quad m = 1,2,3,4 \tag{12}$$

where $\{c, t_0, \theta_m, \phi_m\}$ are real-valued constants, and

$$\Theta = \sum_{m=1}^{4}\theta_m = \arccos\{-\frac{1}{4\delta}\sum_{m,n=1}^{4}\Gamma_{m,n}\}, \tag{13a}$$

$$c^2 = \frac{1}{2\delta \cdot \sin\Theta}, \quad \phi_m = -c^2\{\sum_{n=1}^{4}\Gamma_{m,n} + \delta\cos\Theta\}. \tag{13b,c}$$

[These results hold for $t_0 > t$; for $t > t_0$, one changes the sign of $(t_0 - t)$ in (12), and the sign of $c^2$ in (13b,c).] The four free constants in (12) are $t_0$, and any three of the four $\theta_m$. Then the last $\theta_m$ must be chosen to satisfy (13a). Substitution of (12) into (9) shows that this form of solution is possible if and only if (11) holds. One can also verify by substituting (12) into (8a) and (10) that $\{H, J_1, J_2, J_3\}$ all vanish for any solution in this family.

Next we show that when (11) holds, *all* solutions of (9) blow up in finite time. To do so, we may consider the solution in (9) to be the first term in a Laurent series, near $(t = t_0)$, and seek solutions of (9) in the form (for $m = 1,2,3,4$)

$$B_m(t) = \frac{c \cdot e^{i\theta_m}}{(t_0 - t)^{\frac{1}{2} + i\phi_m}}[1 + \alpha_m(t - t_0) + \beta_m(t - t_0)^2 + ...]. \tag{14}$$

In (14), $\{c, t_0, \theta_m, \phi_m\}$ are defined as above, and $\{\alpha_m, \beta_m,...\}$ are complex numbers. Substituting (14) into (9) and requiring that the complex coefficient of each power of $(t - t_0)$ vanish shows that most of the coefficients in this expansion are fixed, with four exceptions: the real parts of three $\alpha_m$ can be chosen arbitrarily, as can the imaginary part of one $\beta_m$. Thus, the family of solutions in (14) contains eight free, real constants. [For example, one can choose the 8 free constants to be $\{t_0, \theta_1, \theta_2, \theta_3, \text{Re}(\alpha_1), \text{Re}(\alpha_2), \text{Re}(\alpha_3),$

Im($\beta_4$)}.] Therefore the family of solutions in (14) is the *general* solution of (9), so all solutions of (9) with nonzero initial data blow up in finite time, provided only that (11) holds.

Because the solutions in (12) all occur with $\{0 = H = J_1 = J_2 = J_3\}$, it follows that the four new constants in (14) must determine $\{H, J_1, J_2, J_3\}$. One can show that for $m = 1,2,3$,

$$J_m = 2c^2[\text{Re}(\alpha_4) - \text{Re}(\alpha_m)]. \tag{15}$$

Then Im($\beta_4$) determines the value of $H$, but we have found no simple way to write this relation.

It is known that the self-focussing (or "wave collapse") singularity of an NLS-type equation occurs only for a range of $H$ [13]. The singularity in (14) occurs for any (real) values of $\{H, J_1, J_2, J_3\}$, provided only that (11) holds, so the two kinds of singularities differ in this respect. They also differ because spatial structure plays an essential role in a self-focussing singularity, but it plays no role whatsoever here.

It remains to show that the solutions of (9) must be non-singular if (11) is not satisfied, so that (11) is both necessary and sufficient for an explosive instability. Suppose that $|B_4(t)| \to \infty$ as $t \to t_0$. Then it follows from (8a) that all four $|B_m(t)|$ must grow at the same rate. Hence as $t \to t_0$, the dominant terms in (10) are

$$H_1 = -\frac{i}{2} \sum_{m,n=1}^{4} \Gamma_{m,n} |B_4|^4 + O(|B_4|^2), \qquad H_2 = -2i\delta |B_4|^4 \cos(\varphi) + O(|B_4|^3),$$

where $\varphi(t)$ is some (unknown) phase. These dominant terms must balance as $t \to t_0$, so necessarily

$$|\frac{1}{2} \sum_{m,n=1}^{4} \Gamma_{m,n}| \cdot |B_4|^4 = |2\delta|B_4|^4 \cos(\varphi)| \leq 2|\delta| \cdot |B_4|^4 \tag{16}$$

in this limit. Dividing by $|B_4|^4$ shows that no explosive singularity can occur for $|\sum_{m,n=1}^{4} \Gamma_{m,n}| > 4|\delta|$. This completes the proof.

Explosive instabilities due to 3-wave mixing have been known for thirty years [4-8]. To our knowledge, no explosive instability caused by 4-wave mixing has ever been observed in a physical system. The analysis above indicates that it should be possible. As with 3-wave mixing, an explosive instability in a 4-wave system requires a background source of energy, so that all four wave modes can grow in intensity together. And as with 3-wave mixing, the indication that such a background source is available is that all four $\delta_m$ in (6) or (7) have the same sign. One difference between the two processes is that for 4-wave mixing, this agreement in signs of the $\delta_m$ in (7) by itself does not guarantee an explosive instability – the interaction coefficients must also satisfy (11).

This work was funded in part by an NSF-MCTP grant, DMS-0602284.

———————